Title: Inference of ecological interaction networks

Authors: Paola Vera-Licona[a] and Reinhard Laubenbacher [b]

[a] BioMaPS Institute, Rutgers University, Piscataway, New Jersey 08854, USA
[b] Virginia Bioinformatics Institute, Virginia Polytechnic Institute and State University, Blacksburg, Virginia 24061, USA.

Address for correspondence: Paola Vera-Licona, BioMaPS Institute, Rutgers University, Piscataway, New Jersey 08854, USA. Voice: 732-445-3160; fax: 732-445-3168. mveralic@math.rutgers.edu

Abstract
The inference of the interactions between organisms in an ecosystem from observational data is an important problem in ecology. This paper presents a mathematical inference method, originally developed for the inference of biochemical networks in molecular biology, adapted for the inference of networks of ecological interactions. The method is applied to a network of invertebrate families (taxa) in a rice field.

Key Words: ecology, modeling, network inference, network of ecological interactions, reverse engineering, simulation, systems biology.

1. Introduction

*Ecology and Systems Biology*: Biological organisms can be studied at many different levels, from proteins and nucleic acids, to cells, to individuals, to populations and ecosystems, and finally, to the biosphere as a whole. Ecology studies the relationships of organisms with each other and with their physical environment. Although ecology has focused on the higher levels of the organization of life, many sub-disciplines have evolved, integrating many of these different levels of organization (*e.g.* molecular ecology, systems ecology) for a holistic study of organisms. Systems biology is also based on the premise that an understanding of the behavior of biological systems at each level of organization is achieved by careful study of the complex dynamical interactions between the components. It is not surprising then that interesting parallels can be found in problems pertaining to both disciplines, opening the possibility of adapting some mathematical methods developed for the study of biological systems in systems biology to the study of ecological systems. One such problem is that inference (or reverse-engineering) of networks, which is the central focus of this paper.

*Mathematical Modeling and Simulation in Ecology*. Modeling and simulation have proven to be powerful tools in many disciplines of ecology; some examples of this are the Lotka-Volterra competition models (Townsend 2002) and Cushing's nonlinear dynamical models in Population Ecology (Costantino 1995).

There are many ways in which one can classify mathematical models, including whether they give a structural and/or dynamical description of the system. The *structural description* of a network (known also as network topology or static model) provides a description of the elements in the system and which elements interact with each other (given through causal or correlation relationships among entities in the system, for example) where sometimes a weight can be given to such relationships; a common example of these static models is that of food webs. Both in ecology and systems biology, the study of network topology is an important problem. For instance, the search for network topology motifs in (Camacho, Stouffer et al. 2007) is quite similar to the focus on unusually common local topological features in the connectivity graph of gene regulatory networks in (Alon 2007). A *dynamical description* (dynamical models) of a network provides a description of the nature of individual relationships, that is to say, a description of how the systems evolves from a given state; Lotka-Volterra differential equations models are one example of this type of models. Once these models have been developed, they can be simulated and visualized, providing an experimental playground for *in silico* study of scenarios.

*Reverse-engineering as modeling framework.* Within the context of systems biology, important classes of networks are biochemical and gene regulatory networks constructed from time course data such as DNA microarray. Several reverse engineering methods have emerged for the construction of network models from large-scale experimental measurements (Price and Shmulevich 2007) and an understanding of characteristics of the topology of such networks (Alon 2007) as well as the dynamics. See, e.g., (Stolovitzky, Monroe et al. 2007) for a recent survey of reverse engineering algorithms.

*Reverse-engineering methods within the context of ecology.* Modeling and simulation of networks of interactions between species in ecosystems helps to understand common patterns (Abrams et al. 1995), to assess their development (Perez-Espana & Arreguin-Sanchez 2001), to predict the effects of human impacts among natural systems and to prevent biodiversity loss (Dunne et al. 2002), etc. When more than a few species are involved in an ecosystem, the construction of networks of ecological interactions is challenging and mathematical and statistical are powerful tools for their inference using sampled data; see, *e.g.*, (Zhang 2007).

In the subsequent sections we describe one reverse engineering method, which uses time course sampled data for all nodes in the network and returns a static model of the network that matches the observed data. Its characteristics are that it uses a finite number of values for each variable and discrete time. The initial model paradigm was published in (Laubenbacher & Stigler 2004) and has been refined since then, *e.g.,* in (Jarra *et al.* 2007); then we describe how the reverse engineering method can be adapted to rebuild a network of ecological interactions

2. The Method

## 2.1 *Inference method to construct the static network*

The network inference method (Jarrah *et al*. 2007) was developed originally for the inference of biochemical networks, such as gene regulatory networks, from DNA microarray and other molecular data sets. It uses techniques from symbolic computation and algebraic combinatorics, complementing an earlier network inference method using techniques from computational algebra (Laubenbacher & Stigler 2004).

The goal of the inference algorithm is to output one or more most likely static networks for a collection $x_1, \ldots, x_n$ of interacting *ecological units* (species, families, etc), which we will refer to as variables. The state of an *ecological unit* can represent the number of individuals present. That is, each variable $x_i$ takes values in the finite set $X = \{0, 1, 2, \ldots, m\}$. A *static network* in this context consists of a directed graph, whose $n$ nodes are the *ecological units* $x_1, \ldots, x_n$. A directed edge $x_i \rightarrow x_j$ indicates an ecological interaction that can be interpreted as indicating for example, that the survival of $x_j$ depends on $x_i$ The inference algorithm takes as input one or more time courses of observational data. The output is a most likely network structure for $x_1, \ldots, x_n$ that is consistent with the observational data. It is worth emphasizing that the network is constructed from the frequency of the supporting observations alone, in a way that is unbiased by prior knowledge or expected results.

The notion of consistency with observational data makes the assumption that the network of interacting *ecological units* $x_1, \ldots, x_n$ can be viewed as a dynamical system that is described by a function $f: X^n \rightarrow X^n$, which transforms an input state $(s_1, \ldots, s_n)$, $s_i$ in $X$, of the network into an output state $(t_1, \ldots, t_n)$ at the next time step. A directed edge $x_i \rightarrow x_j$ in the static network of this dynamical system $f$ indicates that the value of $x_j$ under application of $f$ depends on the value of $x_i$. A directed graph is a static network consistent with a given time course $\mathbf{s}_1, \ldots, \mathbf{s}_r$ of states in $X^n$, if it is the static network of a function $f: X^n \rightarrow X^n$ that reproduces the time course, that is, $f(\mathbf{s}_i) = \mathbf{s}_{i+1}$ for all $i$.

The algorithm in (Jarrah *et al.* 2007) computes ALL minimal static networks consistent with the given data set. Here, a static network is *minimal* if, whenever an edge is removed, the resulting graph is not anymore a wiring diagram consistent with the data. This process is done one variable at a time, that is, by computing the edges adjacent to individual vertices one at a time, rather than the diagram as a whole. Furthermore, for a given vertex, we use an efficient combinatorial parametrization of the possible edge configurations, rather than by an enumerative method. The next step is to define a probability distribution on the space of minimal possible edge sets for each vertex that permits the selection of a most likely wiring diagram for a given edge. Since the combinatorial description of the space allows the actual computation of this measure on the whole space, this method has the advantage over other methods of choosing the most likely model from the whole space. Heuristic learning methods such as Bayesian network inference typically proceed from a random initial choice of network and find a local minimum of a suitably chosen scoring function. Thus, typically only a small part of the entire space is explored. We illustrate the algorithm with a small example.

Example. Let's assume that we are given the time course data
$\mathbf{s}_1 = (1, 0, 0, 2)$,
$\mathbf{s}_2 = (1, 2, 2, 1)$,
$\mathbf{s}_3 = (0, 2, 1, 1)$,
$\mathbf{s}_4 = (1, 2, 1, 2)$,
$\mathbf{s}_5 = (2, 2, 0, 2)$,
$\mathbf{s}_6 = (0, 1, 1, 2)$

representing the number of individuals of four given species $x_1, \ldots, x_4$, where each of the species has *0, 1,* or *2* individuals. Applying the above algorithm to this data set results in the following output:

$F_1 = \{ \{x_1, x_3\}, \{x_1, x_2, x_4\}, \{x_2, x_3, x_4\} \}$ (1)
$F_2 = \{ \{x_1\}, \{x_2, x_3\}\}$ (2)
$F_3 = \{ \{x_1, x_3\}, \{x_1, x_2, x_4\}, \{x_2, x_3, x_4\} \}$ (3)
$F_4 = \{ \{x_1, x_3\}, \{x_2, x_3\}, \{x_1, x_2, x_4\} \}$ (4)

This output is to be interpreted as follows: For *ecological unit* $x_1$, possible incoming connections ($F_1$) are either $x_1$ and $x_3$ *or* $x_1$, $x_2$, and $x_4$, or $x_2$, $x_3$, and $x_4$. These three sets are minimal, in the sense that the data cannot be explained by choosing a subset of the three possibilities. The other rows are interpreted similarly. One can now "mix and match" possible incoming edge sets for the species and obtain in this way all possible wiring diagrams consistent with the given data set.

With the use of this example, we are able to emphasize that the term *minimal* static network does not imply that a graph with the least amount of edges is constructed; as we can see for the ecological unit $x_2$ there are two choices for its ecological interactions (interaction with $x_1$ or $x_2$ and $x_3$) for which the choice of interactions of $x_2$ with $x_2$ and $x_3$ will return a graph with one more edge than the choice of interaction between $x_2$ and $x_1$.

2.2 Model selection. In order to select a most likely static network among the potentially very large number of possible ones, one defines a probability distribution on the space of all possible static networks for a given data set, which we briefly explain here. The model selection method first scores each of the variables with a formula that is based on the proportion of sets in which it appears. Then it scores sets based on the scores of the variables in them. To be precise, suppose the algorithm outputs the possible variable sets $F_1, \ldots, F_t$, each a subset of the set of all variables $x_1, \ldots, x_n$. For each $s = 1, \ldots, n$, let $Z_s$ be the number of sets $F_i$ that contain $s$ elements. For each $i = 1, \ldots, n$, let $W_i(s)$ be the number of sets with $s$ elements that contain $x_i$. Then define a variable score

$$S(x_i) = \Sigma^n_{s=1} W_i(s)/sZ_s.$$

Using this score, we assign a score $T(F_j)$ to every set $F_j$ in the output by taking the product of the variable scores $S(x_i)$ for all $x_i$ in $F_j$. Normalizing by the sum of all scores $T(F_j)$, we obtain a probability distribution on the set of all $F_j$.

With the help of this probability distribution, we can now choose the set(s) with highest probability as the most likely static networks. In the case of a tie, a final selection will have to be made based on biological considerations.

Example. We apply this measure to the data set in the example above, focusing on specie $x_1$. Then

$$F_1 = \{x_1, x_3\},\ F_2 = \{x_1, x_2, x_4\},\ F_3 = \{x_2, x_3, x_4\},$$

and we obtain the following variable scores:

$$S(x_1) = 1/1 + 1/(2\cdot3) = 7/6,\ S(x_2) = 2/(2\cdot3) = 1/3,$$
$$S(x_3) = 1/1 + 1/(2\cdot3) = 7/6,\ S(x_4) = 2/(2\cdot3) = 1/3.$$

Finally, the sets are scored as follows:
$T(F_1) = (7/6)(7/6) = 49/36$, $T(F_2) = (7/6)(1/3)(1/3) = 7/54$, $T(F_3) = (1/3)(7/6)(1/3) = 7/54$.
Based on these scores, we choose $F_1$ as the most likely set of incoming connections for $x_1$. Carrying out a similar computation for the other two variables results in a complete, most likely static network for the given data set.

## 3. An Application

In order to illustrate an application of the reverse engineering method introduced, we will use published results on a network of ecological interactions in (Zhang 2007) in order to evaluate our method's performance.

In Zhang's, a network inference method is introduced and validated with the use of a set of invertebrates data sampled in a rice field (Zhang *et al.* 2004). There a total of 75 invertebrates families (or taxa) and 60 samples are considered.

For the purposes of the present paper we focus on a sub network from Zhang's, corresponding to the invertebrate family *Culicidae* (see Zhang 2007: fig. 2), which consists of *9* invertebrate families that according to Zhang, *16* ecological interactions among them exist; we restrict to only *20* (of the *60*) samples for these *9* taxa (see Table 1). For model selection, we considered the interactions with scores (as described in the previous section) above *.50*.

The network obtained with (Jarrah *et al.* 2007) method, with also *16* ecological interactions, is depicted in (Figure 1). All the biological families are linked to the rice field, but they do not correspond to the category of ecological interactions. With the use of only a third of the data used to build the network in (Zhang 2007), we found *12* of the *16* expected interactions. One of the missing interactions corresponds to the one between the biological families *Culicidae* and *Dryinidae*, which appear as one of the highest scored interactions but not above *.50*; instead, we obtain the ecological interaction between the families *Dryinidae* and *Carabidae*; one possible explanation for the existence of this interaction is that, *Carabidae* (as part of the Coleopteran insects) may depict with the *Dryinidae* a parasite-parasitoid interaction.

On the other hand, we observe that the total number of ecological interactions for this network agrees with that of Zhang's (total of *16*), and, therefore, the ratio between the total number of interactions and the number of biological families is preserved (*16/8 = 2*).

## 4. Discussion

The problem of inferring a network of interactions in a biological system from sampled data appears in several different, apparently disparate, contexts. This paper focuses on two such contexts, the inference of ecological interactions in an ecosystem and the inference of biochemical networks in systems biology. We have shown that a method designed for this purpose in one field can be applied profitably in the other one. Using a data set consisting of a time course of observations of several different species in a common context we have inferred a network of interactions between these species, represented by a graph whose nodes are the species and whose edges represent interactions. In molecular biology, such a representation would be called a *wiring diagram.*

*Contribution of mathematical inference methods.* One advantage of the inference method presented here is that it uses a sophisticated mathematical encoding of the entire space of possible wiring diagrams consistent with the data. Based on the selection criteria chosen, it then selects exactly ALL static networks that fit these criteria. In contrast, statistical methods for network inference, such as Bayesian networks, find an optimal solution through a heuristic search.

As in molecular biology, ecological networks typically are dynamical systems that change over time. This is reflected in the fact that many mathematical models in ecology are dynamic, typically represented by systems of differential or difference equations. It would therefore be desirable to have a method available that infers not only a wiring diagram but a dynamic description of the system. Another advantage of the method presented here is that there are methods closely related to the one presented here which are able to do just that (Laubenbacher & Stigler 2004; Dimitrova, Jarrah *et al.* 2007). However, typically more data are required to be able to infer accurate models. Furthermore, in order to infer the causal relationships among dynamic variables it is very useful to be able to perturb the system in different ways. In molecular biology perturbations are typically done by “knocking out” genes, in gene regulatory networks or interfering in some other ways with the action mechanisms of individual systems variables. In ecosystems this may be more difficult to accomplish.

In addition to the method presented here, there are several other inference methods available for molecular network. A study of their usefulness to help solve problems in ecology would be of interest. For instance, several such methods allow the introduction of prior biological knowledge into the inference process (Tsai and Wang 2005; Cosentino *et al.* 2007), thereby improving algorithm performance.

**Table 1.** Sampling data of rice invertebrates as shown in (Zhang 2007). Highlighted is the data we considered for 9 different families (taxa) for the *Culicidae* subnetwork; these data (20 samples) represent *30%* of the data originally used in (Zhang 2007) to infer such subnetwork.

| | **1** | **2** | **3** | **4** | **5** | **6** | **7** | **8** | **9** | **10** | **11** | **12** | **13** | **14** | **15** | **16** | **17** | **18** | **19** | **20** |
|---|---|---|---|---|---|---|---|---|---|---|---|---|---|---|---|---|---|---|---|---|
| *Elateridae* | 2 | 0 | 4 | 0 | 0 | 1 | 0 | 0 | 2 | 1 | 0 | 0 | 0 | 3 | 0 | 1 | 0 | 0 | 1 | 0 |
| *Culicidae* | 3 | 4 | 0 | 0 | 1 | 0 | 6 | 4 | 3 | 0 | 5 | 5 | 5 | 1 | 0 | 2 | 4 | 1 | 10 | 0 |
| *Coenagrionidae* | 0 | 0 | 0 | 0 | 0 | 0 | 0 | 1 | 2 | 0 | 0 | 0 | 0 | 0 | 0 | 0 | 2 | 0 | 0 | 0 |
| *Aleyrodidae* | 0 | 0 | 0 | 0 | 0 | 0 | 0 | 1 | 0 | 0 | 0 | 0 | 0 | 0 | 0 | 0 | 0 | 0 | 0 | 0 |
| *Mymaridae* | 0 | 0 | 0 | 1 | 1 | 0 | 0 | 0 | 1 | 0 | 0 | 0 | 0 | 0 | 1 | 0 | 0 | 0 | 1 | 0 |
| *Chloropidae* | 0 | 0 | 0 | 1 | 0 | 0 | 0 | 0 | 0 | 0 | 0 | 0 | 0 | 0 | 1 | 0 | 0 | 1 | 1 | 0 |
| *Grylliade* | 1 | 0 | 4 | 5 | 2 | 6 | 0 | 2 | 4 | 2 | 4 | 3 | 1 | 0 | 2 | 3 | 5 | 0 | 0 | 2 |
| *Araneidae* | 1 | 0 | 0 | 0 | 0 | 0 | 0 | 0 | 0 | 0 | 0 | 0 | 0 | 1 | 0 | 0 | 0 | 0 | 1 | 0 |
| *Theridiidae* | 0 | 0 | 0 | 0 | 0 | 0 | 0 | 0 | 0 | 0 | 0 | 0 | 0 | 1 | 0 | 0 | 0 | 0 | 0 | 0 |
| *Baetidae* | 0 | 0 | 0 | 0 | 0 | 0 | 0 | 0 | 0 | 0 | 0 | 0 | 0 | 0 | 0 | 0 | 1 | 0 | 1 | 0 |
| *Hydrophilidae* | 0 | 0 | 0 | 0 | 0 | 0 | 0 | 0 | 0 | 0 | 0 | 0 | 0 | 0 | 1 | 0 | 0 | 0 | 0 | 0 |
| *Blattellidae* | 0 | 0 | 2 | 0 | 0 | 0 | 0 | 0 | 0 | 0 | 0 | 0 | 0 | 0 | 1 | 0 | 0 | 0 | 1 | 0 |
| *Braconidae* | 0 | 0 | 0 | 0 | 0 | 0 | 0 | 0 | 0 | 0 | 0 | 0 | 0 | 0 | 1 | 0 | 0 | 0 | 1 | 0 |
| *Cicadellidae* | 2 | 1 | 0 | 5 | 0 | 2 | 1 | 0 | 5 | 1 | 1 | 1 | 0 | 1 | 3 | 1 | 0 | 1 | 2 | 0 |
| *Miridae* | 3 | 2 | 0 | 1 | 0 | 0 | 0 | 1 | 0 | 0 | 1 | 0 | 0 | 0 | 0 | 1 | 0 | 0 | 2 | 0 |
| *Tettigoniidae* | 1 | 0 | 0 | 0 | 0 | 1 | 0 | 0 | 0 | 0 | 0 | 1 | 0 | 0 | 0 | 1 | 1 | 0 | 0 | 0 |
| *Linyphiidae* | 2 | 1 | 1 | 0 | 3 | 0 | 3 | 1 | 0 | 4 | 1 | 1 | 3 | 0 | 3 | 2 | 2 | 1 | 2 | 4 |
| *Ceratopogonidae* | 0 | 2 | 0 | 0 | 1 | 0 | 1 | 0 | 0 | 0 | 0 | 0 | 1 | 0 | 0 | 0 | 0 | 0 | 1 | 1 |
| *Chironomidae* | 0 | 2 | 0 | 0 | 0 | 3 | 1 | 1 | 2 | 2 | 0 | 1 | 2 | 0 | 1 | 0 | 0 | 0 | 1 | 4 |
| *Encyrtidae* | 0 | 0 | 0 | 0 | 0 | 0 | 0 | 0 | 0 | 0 | 0 | 0 | 0 | 0 | 0 | 1 | 0 | 0 | 0 | 0 |
| *Cunaxidae* | 1 | 0 | 1 | 0 | 0 | 0 | 0 | 1 | 0 | 0 | 0 | 0 | 0 | 2 | 0 | 0 | 1 | 0 | 0 | 0 |
| *Drosophilidae* | 0 | 0 | 0 | 1 | 0 | 0 | 0 | 0 | 0 | 1 | 0 | 0 | 1 | 0 | 1 | 0 | 0 | 0 | 0 | 0 |
| *Dryinidae* | 2 | 0 | 0 | 0 | 0 | 0 | 0 | 1 | 0 | 0 | 0 | 0 | 0 | 0 | 0 | 0 | 0 | 0 | 0 | 0 |
| *Tetragnathidae* | 2 | 0 | 0 | 2 | 1 | 3 | 4 | 3 | 0 | 1 | 4 | 1 | 0 | 1 | 0 | 1 | 1 | 3 | 3 | 2 |
| *Dytiscidae* | 0 | 0 | 0 | 1 | 0 | 1 | 0 | 0 | 0 | 0 | 0 | 0 | 1 | 0 | 0 | 0 | 0 | 0 | 0 | 0 |
| *Carabidae* | 1 | 0 | 0 | 0 | 0 | 0 | 1 | 0 | 1 | 2 | 1 | 0 | 0 | 0 | 0 | 0 | 1 | 0 | 1 | 0 |
| *Entomobyidae* | 2 | 1 | 2 | 2 | 0 | 0 | 1 | 2 | 0 | 0 | 1 | 0 | 0 | 0 | 3 | 7 | 0 | 0 | 0 | 1 |
| *Hydrometridae* | 0 | 0 | 0 | 0 | 0 | 2 | 0 | 0 | 0 | 0 | 0 | 0 | 0 | 0 | 1 | 1 | 0 | 0 | 0 | 0 |
| *Hydraenidae* | 0 | 0 | 0 | 0 | 0 | 1 | 0 | 0 | 0 | 1 | 0 | 0 | 0 | 0 | 0 | 0 | 0 | 0 | 2 | 0 |

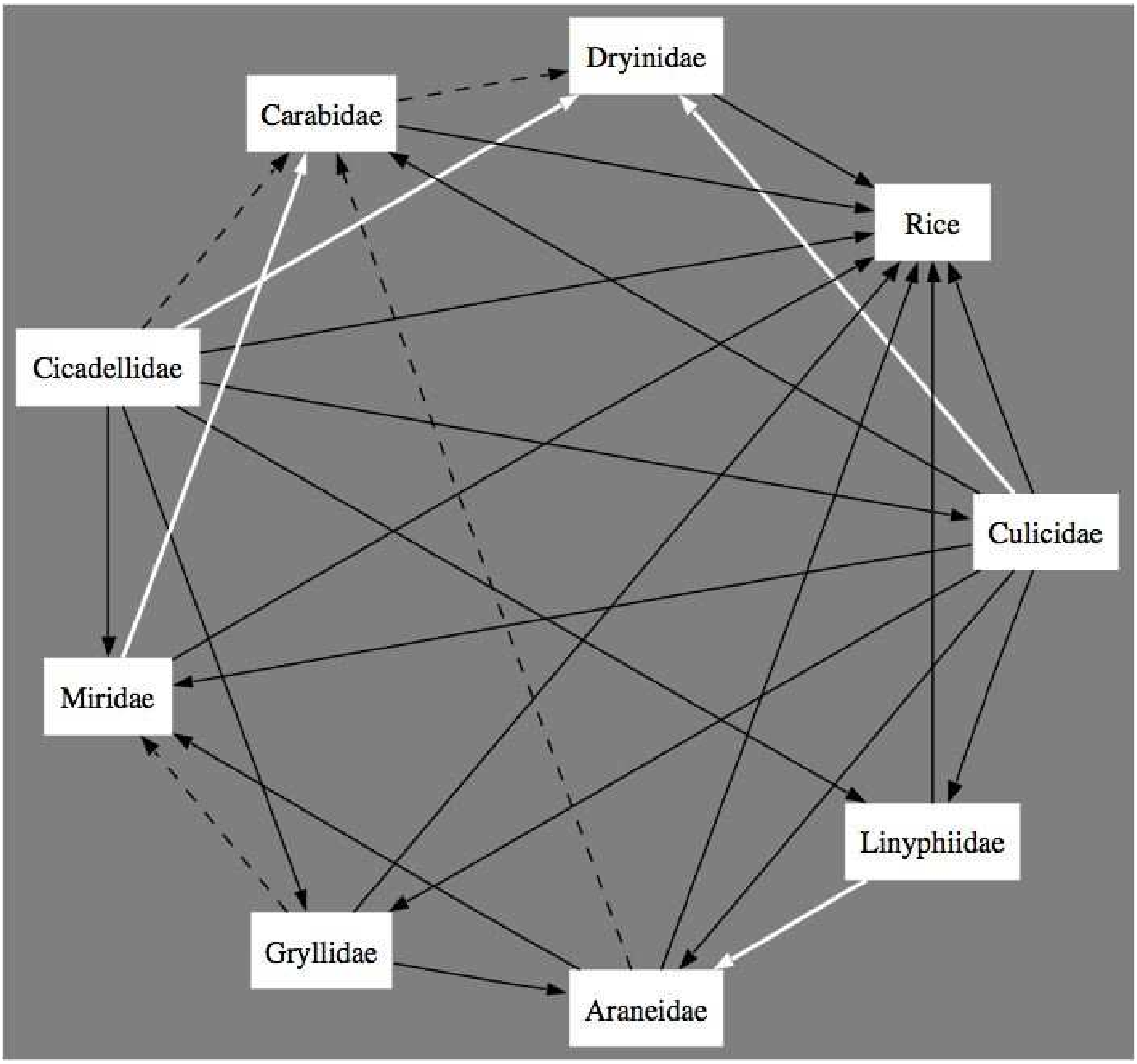

**Figure 1.** Network of ecological interactions inferred for the *Culicidae* subnetwork from Table 1. Filled edges represent edges that are found in both, the subnetwork built in (Zhang 2007) and **(**Jarrah *et al.* 2007); dashed edges represent the relationships found in (Jarrah *et al.* 2007) but not in (Zhang 2007); white edges represent edges missing in **(**Jarrah *et al.* 2007) from those found in (Zhang 2007).